\documentstyle[12pt]{article}

\def\be{\begin{equation}}
\def\ee{\end{equation}}
\def\bea{\begin{eqnarray}}
\def\eea{\end{eqnarray}}

\def\ni{\noindent}

\begin{document}
\begin{center}
\Large
{\bf We have time because we shall never know}
\end{center}
\vspace{.3in}
\normalsize
\begin{center}
Reza Tavakol$^1$ and Edward Anderson$^2$ \\
\end{center}
\normalsize
\vspace{.1cm}
\begin{center}
$^1${\em Astronomy Unit, School of Physics and Astronomy, \\
Queen Mary University of London,\\
Mile End Road, \\ London. E1 4NS. UK \\} 
\vspace{0.4cm}
$^2${\em DAMTP, \\
Cambridge University,\\
Wilberforce Road, \\ Cambridge. CB3 0WA. UK \\}
\date{The date}
\end{center}
\vspace{1cm}

%--------------------------------------------------------
\begin{abstract}
%--------------------------------------------------------

We argue against current proposals concerning the non-existence of time.
We point out that a large number of these proposals rely,
at least implicitly, on the assumption of `closure' (or `partial closure') of the laws of Physics.
I.e. the assumption that laws of Physics as they are known today are
either complete (and hence closed) or that they possess features that
a hypothetical future `complete' theory must share
(and hence are partially closed).
Given that the assumption of closure of laws
of Physics can never be verified operationally, it
cannot justifiably be used to support the claim for non-existence of time.

Some approaches against time are `timeless'
at the primary level for the universe as a whole.
In these approaches time arises at a secondary level, mostly
in the sense of `time being abstracted from change'.
On the other hand, there are other approaches that 
deny the existence of time altogether.
We argue that metaphysical arguments of this type 
-- similar to those based on closure --
by implicitly implying the absence of history,
are by their nature circular.

%--------------------------------------------------------
\end{abstract}
%--------------------------------------------------------

\newpage

%--------------------------------------------------------
\section{Introduction}
%--------------------------------------------------------

It has become rather fashionable to deny the existence of time 
\cite{PW83, NSI, B94I, B94II, Page, Rec-In-Hist, H99, Rovelli, H03}.
The idea of ultimate reality being timeless has a 
long prehistory among our ancestors \cite{Whitrow-89}.

Over recent years there has been a great deal of renewed interest 
in the question of time in both Physics and Philosophy.
These considerations have been varied and numerous, 
so to maintain clarity it is useful to adopt a clear terminology.
In the following we shall employ {\it time world-views} to indicate 
metaphysical positions regarding time (such as the `block', `Broad' and
'solipsist' world-views discussed below).  
This is in order to keep these clearly
distinct from the status of time in {\it physical paradigms}\footnote{These 
are families of physical theories which each have their own notion of time: 
Newtonian Physics, Special Relativity (SR), General Relativity (GR), 
or various more speculative Quantum Gravity programs.} 
in which these world-views are realized to some extent

Two key sets of questions considered in the literature
in connection with the problem of time are: (i) those concerning the 
existence or non-existence of time and (ii) those relating to the
mismatch between the notions of time in different theories.

Many assertions concerning the non-existence of time find their 
justifications in the context of recent attempts at developing a theory of Quantum Gravity.  
In approaches to Quantum Gravity, one is often interested in bridging between theories. 
The difficulties then often stem from time {\sl not} having the same meaning within each 
theory\footnote{So for example, ordinary Quantum Mechanics 
(QM) largely inherits its notion of time from Newtonian Physics, or from SR in the case of
Quantum Field Theory (QFT), neither of which are compatible with
GR's notion of time.} 
in its standard formulation. 
Moreover, a given theory can often be interpreted with 
multiple world-views concerning time. 
Thus one may attempt to change the formulation of one or both of the theories
such that the new world-views concerning time match.
Many approaches to the problem of time are indeed of this nature 
\cite{Kuchar92, I93, Anderson}. 
Thus for example if temporal issues cause incompatibility
between two theories, 
one might attempt to reformulate them 
such that time is absent in both, 
or seek hidden times in both theories that are akin. 
One could also attempt to reformulate both QM and GR as
histories theories so as to bridge the gap.
Alternatively, one could try to find matching emergent
times in the theories one is considering.

Some time world-views considered in the literature are as follows: 
{\it Timeless solipsism} (the most extreme of such positions,
going back to Parmenides and Zeno), according to which  the `present now' 
exists while the past or future do not. Thus this position only recognises {\it being}
as opposed to {\it becoming}. In contrast, there are other positions which 
recognise becoming to different extents by ascribing
reality to time (such as for e.g. Heraclitus,
who considered time itself to flow),
or at least to a subset of properties that are intuitively `temporal'.
For example, according to the {\it block universe}
world-view \cite{Earman-Block, Capetown-Proc}, which 
originated in the 19th century \cite{Jammerteneity},
not only the present and the past but also the future 
(and hence the entire `spacetime block') is already there. 
Thus in this case there is no sense in which we can
talk about a flow of time. This position is to be contrasted with  
{\it Broad's world-view}\footnote{This is also known as
the {\it growing block}, though we do not use this term so as not to 
give the impression that it is  a subcase of the block world-view,
but rather a separate world-view in its own right.} \cite{Broad, Earman-Block}, 
according to which the present and the past exist
while the future unfolds in time.

Finally, a position which goes back to Leibniz \cite{L, Whitrow}
treats time for the universe as a whole as meaningless at the primary level. 
This may appear to be compatible with solipsism, but since it still allows
change at the primary level, one could say with 
Mach \cite{M} that time in this case `is to be abstracted from change'.
\vskip 0.05in
In this article we consider some recent 
arguments for the non-existence of time.
In particular we shall consider:

\begin{itemize} 

\item  Block universe type arguments \cite{Earman-Block, Capetown-Proc}.

\item Arguments from so-called timeless Quantum Gravity programs, such as:

\begin{itemize}

\item[(i)] The apparently frozen form of the Wheeler-DeWitt equation \cite{Kuchar92, I93, Peres}.

\item[(ii)] Page's argument \cite{Page}. 

\item[(iii)] Bubble Chamber type arguments in Records Theory \cite{Bell, B94II, H03}.

\item[(iv)] The emergent semi-classical time approach \cite{HallHaw, Kiefer} to e.g. Quantum Cosmology. 

\end{itemize}

\end{itemize}

%\mbox{ } 

\noindent Argument (ii) -- and some arguments of the types (i) and (iii) -- 
are used in various implementations of the solipsist world-view.
Such arguments are metaphysical in nature and 
by implicitly denying history, they 
are circular and cannot be
used as arguments against the existence of time.\footnote{To avoid confusion,
note that in this article we use `history' in the ordinary use of the term, rather 
than its technical use in `histories approach' to Quantum Theory.}
We shall therefore concentrate on the other arguments 
concerning the non-existence of time
and point out that many of these arguments rely, at least implicitly,
on the assumption of `closure' (or `partial closure') of the laws of Physics,
assumptions that we argue cannot be made justifiably.

The article proceeds (in Section \ref{Closure}) by considering 
notions of `closure' and `partial closure' of laws of Physics 
along with discussion of why these notions are defective.
In Section \ref{particular} we consider some of the main arguments 
in the literature for the non-existence of time 
and show how they employ closure type arguments. 
Interestingly, closure type arguments also implicitly appear in contrary settings, 
in arguments for the existence of time. 
This raises questions about the uniqueness of such notions of time.
We discuss some examples of this in Section \ref{existence}.

%--------------------------------------------------------
\section{Assumption of closure of laws of Physics}\label{Closure}
%--------------------------------------------------------

Two key assumptions that are often made regarding the laws of Physics, 
at least implicitly, in arguments concerning the non-existence of time are as follows.

\begin{enumerate}

\item[I.] {\bf Closure.} 
The assumption that laws of Physics as they are currently known are complete. 
%
% 'known at present' -> 'currently known' 
%
Furthermore, given that laws of Physics commonly take the form of differential equations, 
an additional related assumption, often made implicitly and rarely discussed, 
is that of complete know-ability (in practice) of the underlying initial/boundary conditions.

\item[II.] {\bf Partial closure.} The assumption that  there exist 
ingredients/elements of laws of Physics as they are known at present
that will remain unchanged in all future evolutions of these laws,
and importantly will be shared by any hypothetical future `complete' theory.
In this sense they can be said to be partially 
closed.\footnote{Clearly we are not talking about
stability of approximate predictions of laws of Physics, which enter 
into correspondence  principles, such as for example the need for future 
theories to agree with Newtonian or general relativistic predictions to 
appropriate degrees of approximation.
}
 
\end{enumerate}

An example of arguments for non-existence of time of type (I) 
arises in the so called `block universe' world-view
(see Sec \ref{block}), while an example of type (II) 
arises due to the difficulties that seem to appear
in incorporating time in current formulations of Quantum Gravity.
In particular, the fact that no explicit time occurs in the Wheeler-DeWitt equation (WDE)
(see Sec. \ref{wd}).
\\

In the next section we shall argue that these assumptions are not 
justifiable by summarising a number of their fundamental problems.

%------------------------------------------------------------
\subsection{Problems with the idea of closure}
%------------------------------------------------------------

A key question in the development of Science is whether we can have closure 
(or partial closure) in scientific laws at any given point in history? 
In particular, is it ever possible to justifiably argue that our present 
understanding of the laws of Physics governing the evolution of the Universe is complete? 
\\

There are a number of reasons why the answer to this question is likely to be negative.

\begin{itemize}

\item {\bf Scientific methodology.}
Given the nature of scientific explanation, 
theories must remain open to future tests and hence possible modifications. 
The idea of closure clearly closes this possibility.
Furthermore, the assumption of closure remains 
operationally unverifiable at any point in history,
and as a result can never be established with certainty in practice.

\item {\bf Possible Ceteris Paribus nature of laws of Physics.} 
Even if we assume that we know particular laws of Physics completely,
it is always possible that we have left out some others. 
This amounts to the possibility that scientific laws are in fact
{\it Ceteris Paribus laws}  \cite{Cartwright}, 
in the sense that their predictions are always subject to other possible 
(as yet unknown) laws being kept constant or ignored altogether. 
Were this the case, it would compromise the long term predictions 
of the past and the future.
See, however, the discussion in \cite{HK-RT-1993} regarding the relation between
the stability of the laws of Physics, at any given point in history,
and the extent to which their predictions may be effected 
by their potential Ceteris Paribus nature.

\end{itemize}

In addition to these two potentially unavoidable sources of 
ignorance we also have:

\begin{itemize}

\item {\bf Partial ignorance about initial/boundary conditions}.
In cosmological settings it is extremely difficult, 
if not impossible in practice,
to determine precisely the
initial/boundary conditions necessary to determine the evolution of the
Universe into the future (or indeed the past).
This ignorance also plays a key role in bringing about the 
additional unpredictability due to the possibility of chaos,
discussed below.

\item {\bf The possibility of chaos}.
Since most known laws of Physics are nonlinear in nature, the possibility
of chaos (and sensitive dependence on initial conditions) would be very likely
in the Universe \cite{Tavakol78, Ellis-2012}.
This implies that even if the laws of Physics where known completely, 
precise future predictability would
require the knowledge of initial states infinitely accurately,
which is clearly not possible in practice, specially in view of unavoidable
ignorance regarding initial/boundary conditions discussed above.

\end{itemize}

%--------------------------------------------------------------------
\section{Particular arguments for non-existence of time}
\label{particular}
%--------------------------------------------------------------------

In the previous section we summarised some fundamental problems with the
assumption of closure (or partial closure) in laws of Physics.
It is instructive to ask how these assumptions
in fact enter specific arguments for the non-existence of time. 
We shall do this by considering a number of examples.

%------------------------------------
\subsection{The `block universe' type arguments}
\label{block}
%------------------------------------

According to this world-view, not only the present and the past
but also the future (and hence the entire `spacetime block')
is knowable completely and in that sense already there.
On this basis, 
it is argued that there is no sense in which we can
talk about a flow of time. 
A  key underpinning assumption in this world-view is
that of total know-ability of the past and future (as well as present).
Clearly such knowledge, specially about the future (but also about the past),
can only be obtained if we were in possession of the complete laws of Physics.
Thus it involves an implicit assumption of closure of laws of Physics.
In addition we need to assume that we know the corresponding 
initial/boundary conditions precisely.  
This is crucial, particularly given the extreme difficulties in reconstructing 
such information as well as the nonlinear nature of most laws of Physics, 
and hence the possibility of chaos discussed above.
\vskip 0.05in

In addition, to have complete access to the past or the future we need to
make the further assumption of non-Ceteris Paribus nature of these laws,
which is clearly not justified a priori.
Furthermore, there is a fundamental question regarding the observability of
such a `block'; as there can be no causally connected observers `outside the block'! 
Finally, the fact that blocks are pre-existing renders 
the block world-view particularly limited as regards Geometrodynamics
and emergent spacetime approaches which grew from Wheeler's
arguments against spacetime primality \cite{Battelle}.

%-------------------------------------------------
\subsection{Argument based on frozen equations}
\label{wd}
%-------------------------------------------------

Some of the key historical arguments concerning the non-existence of time
are based on the difficulties in incorporation of time in some 
attempts at formulating Quantum Gravity and the fact that some 
of these approaches lead to so called frozen equations which do
not explicitly contain time. 
E.g. the WDE for the wave function of the Universe,  
and the way it has been taken as an argument for the non-existence of time.
This equation arises from taking the GR Hamiltonian constraint 
${\cal H}$ and canonically quantising it by promoting 
a subalgebraic structure of 
the phase space variables
and the constraints to operators to give the WDE:

$$\widehat{\cal H}\Psi = 0.$$
Here $\Psi$ is the wave function of the Universe.
The explicit absence of time in this equation has been taken as an 
argument for non-existence of time. 

A number of arguments have been made against this interpretation.
To begin with, there are criticisms to the WDE approach itself \cite{Kuchar92, I93, Peres}. 
These include:
(i) whether it is meaningful to associate a wave function with the entire Universe, 
and (ii) whether such a global concept can match our everyday 
local experiences? 

Another important question concerns the extent to which 
equations such as the WDE can be treated as complete. 
Clearly not all matter species may have been taken into account \cite{HP11}. 
Furthermore, time may be more apparent in one set of canonical variables 
than others, e.g. passing from the 3-metric variables of Geometrodynamics
to a set of variables including York hidden time \cite{York73, Kuchar92, I93}.

Furthermore, in using the WDE as a basis for arguments against time,
a crucial assumption, at least implicitly, is that even though
we do not have a full formulation of Quantum Gravity at
present, the timeless feature of the WDE 
is prototypical and would be shared with any hypothetical 
future `complete' theory. This amounts to the assumption of partial closure.
The important question is to what extent do we expect this
assumption to be valid?
To answer this question, we need to ask which are the 
features from which the WDE 
inherits its time-independence and whether these are features that will be 
shared with a hypothetical final theory of Quantum Gravity.
An important ingredient of the WDE that has
been held as important in this regard is background independence.

So the questions are (i) does background independence
necessarily imply the WDE and (ii) to what extent is background independence
expected to be an ingredient of any `final' theory of Quantum Gravity?

The answer to these questions are not known at present. 
It is however useful to note that background independence is a feature shared by 
Geometrodynamics and Loop Quantum Gravity, as well as 
canonical Supergravity and M-Theory (but not the perturbative part of String Theory).
Nevertheless, we recall that regarding (i) Supergravity, for example, has 
sufficiently different realisations of background independence that might 
render frozen equations less 
significant.\footnote{Note that Background Independence has many aspects, in fact in direct
correspondence to the {\sl consequent} Problem of Time facets \cite{Anderson}.}
In this way, the Problem of Time can openly be laid out as the quantum gravitational
%
% 'arena' deleted
%
manifestation of the absolute versus relational motion debate.
Concerning (ii) we should note that if for example M-Theory turns out to be 
holographic in nature, it could in some sense be background dependent: 
on a `screen at infinity, upon which there are privileged frames'. 

Finally, as a small aside regarding (i), not all background independent
theories of gravity possess a frozen wave equation.
In particular, \cite{Boulware} contains a counterexample amongst the higher derivative theories of gravity. 

%------------------------------------ 
\subsection{Naive Schr\"{o}dinger interpretation}
%------------------------------------

Here we have in mind `Schr\"{o}dinger' in the sense of 
involving a Schr\"{o}dinger inner product \cite{NSI} 
for computing timeless relative
probabilities; `interpretation' in the QM sense of involving an alternative to 
the standard Copenhagen interpretation
and `naive' in the sense of not involving the GR constraints, 
which limits this approach's applicability \cite{Kuchar92}. 
This approach involves asking questions of a subsystem or Universe with no
reference to 'when', 'how long for' or 'whether that state is attained in
permanence at some point'. 
This is timeless in the sense of making no reference to time, rather than in the sense of restriction to a
single instant, and is not expected to cover all physically meaningful propositions or investigations.
Nevertheless, some elements of 
this approach resurface within some of the 
approaches below, justifying its mention here.

%------------------------------------ 
\subsection{Bubble chamber arguments}
%------------------------------------

Bubble chamber $\alpha$-particle tracks can be explained in terms of a
%
% 'a' added
%
time-independent Schr\"{o}dinger equation and can therefore be
treated as `timeless' \cite{Mott}.
This has been extended to provide various timeless approaches to Quantum Cosmology.
In \cite{B94II}, Barbour draws upon this fact to argue for formation of
``{\it time capsules}" in space, leading to a type of timeless Records Theory; on the other hand, 
\cite{H03}, Halliwell considers such tracks {\sl in configuration space}.

Clearly one can also give an explanation of bubble chamber 
tracks within a world-view in which time is presupposed.
However, bubble chambers are carefully selected environments for revealing tracks.
In general, records would be expected to be poorer \cite{H99};
perhaps far poorer, along the lines of the Joos--Zeh model -- of a dust particle
decohereing due to the CMB photons \cite{JZ} --
being a far more likely situation to encounter in nature than a bubble chamber.
In such a situation, records would be exceedingly diffuse due to the 
information being dispersed by 
the CMB photons and ending up spread over cosmological space.

%------------------------------------ 
\subsection{Page type argument}
%------------------------------------ 

The conditional probabilities interpretation \cite{PW83} succeeds in supplanting being-at-a-time 
by timeless correlations between the configurations of the studied subsystem and of the clock used. 
Page subsequently considered whether becoming could be supplanted as well. 
His approach \cite{PAOT, Page} to this question involves 
correlations within a single present instant configuration 
which contains memories of what might otherwise be regarded as a sequence of `previous configurations'.
One might view such configurations as e.g. researchers with data sets 
who remember how they set up the experiment that the data came from 
(controlled initial conditions, and so on).
In this scheme, it is not the past instant that is involved, 
but rather its memory or record at the present instant, 
alongside the configuration of what would usually be regarded 
as the `subsystem under study' itself.
This is an `in-principle' scheme, meaning that no claims are made 
about being able to compute these correlations in practise.
However, Hartle's IGUS (information gathering and utilising subsystem) concept \cite{IGUS}
may come to permit such calculations.

Page's argument does not carry over to the GR context of the whole universe.  
It can in principle only be implemented 
for the past events that one can have access to,
i.e. along ones past light cone.  In addition to the difficulties in 
accomplishing this in practice,
we have the important added difficulty of how to uniquely reconstruct the
structure/inhomogeneity of the Universe off the light cone.
Furthermore, and importantly, this only applies to past events and not future ones.
The crucial point is that future records do not exist.
It is this inability to precisely reconstruct the totality of the past records,
as well those in the future, which we call time!
Importantly also, this approach cannot be applied to the Universe as a whole. 
In that case the observer
needs to be at a point in the history of the
Universe whose past light cone covers the entire Universe.  
At any other points in the history of the Universe the records would not be complete.
\vskip 0.05in
Before ending this section, we should also note that
timeless approaches -- such as the na\"{\i}ve Schr\"{o}dinger interpretation \cite{NSI}, 
programs inspired by the bubble chamber model arena,
conditional probabilities interpretation,
and Page's approach 
-- are pretty widely tied to nonstandard interpretations of QM. 
Such approaches either invoke the WDE and so inherit some of its problems, 
or do not, thus risking the alternative problem of being incompatible with WDE \cite{Kuchar92}. 

%-----------------------------------------------------
\section{Closure arguments in Broad approaches containing time}
\label{existence}
%-----------------------------------------------------

Interestingly the assumptions of closure or partial closure are not only used in 
arguments for the non-existence of time, but also in some 
arguments {\it for} the existence of time.
An important example\footnote{Almost all background independent 
Quantum Gravity programs can be taken to reside within this.}
is that of Broad's world-view,
which assumes the present and the past exist,
while the future unfolds in time. 
In these cases the implicit employment of closure or partial closure 
implies a non-unique characterisation of the future 
(and past) dynamics, and hence of time.
However, as opposed to the case of block universe, in this case the presence of 
closure {\it does not} undermine the presence of time, but rather 
makes it non-unique or approximate.

In this Section we consider various
programs which can be argued to sit within Broad's world-view.

%------------------------------------
\subsection{Classical GR}
%------------------------------------
Broad's world-view can be argued to apply to two 
commonly used formulations of GR. 

Firstly, approaches centring on spacetime causality \cite{Wald-HE}.
Observationally this involves study of the past light cone. 
Secondly, approaches based on Geometrodynamics.
From a practical perspective, these are extensively used in numerical GR.
Although the connection of these approaches to Broad's world-view is clear, 
the modern considerations of it in GR only started long 
after the Wheeler school's emphasis on Geometrodynamics in the 1960's. 

Additionally there is a Machian \cite{B94I} interpretation of Geometrodynamics 
following from a Baierlein--Sharp--Wheeler type action \cite{BSW,AM13}.
The emergent time abstracted from change here coincides with GR proper time.

%------------------------------------
\subsection{ Causal Sets program}
%------------------------------------

The underlying structure here \cite{Sorkin} consists of discrete 
objects (the so called `spacetime atoms'),
subject to causal ordering relations under which the Universe model is 
a partially ordered set.
To be clear, most modelling to date have been at the 
{\sl classical} rather than quantum level: 
a classical world governed by causality, and yet 
with less structural assumptions than those
made in SR or GR. 

Dowker \cite{Dowker} has recently argued that this program lies within Broad's world-view; 
where in this case `the block grows' sequentially due to the birth of spacetime atoms.
In our opinion, the connection to Broad's world-view is unsurprising, since that world-view applies to 
the standard spacetime causality approach to GR, which the Causal Sets program shares some 
common foundations with.

%-------------------------------------------
\subsection{Semi-classical GR}
%-------------------------------------------

Finally we consider semi-classical 
GR in a canonical sense and thus closely tied to Geometrodynamics.
This involves interpreting the WDE within a regime in which a semi-classical (WKB) time emerges.
Furthermore, this also admits a Machian interpretation, in terms of time being abstracted from 
semi-classical change which distinguishes it from the classical emergent Machian time by having
a quantum input \cite{Anderson}. 

%------------------------------------
\section{Conclusion}\label{conclusion}
%------------------------------------

We have argued against recent proposals regarding the non-existence of time.
These proposals  fall into two main categories. 
Firstly those arguments that assume, at least implicitly,
the closure of partial closure of the laws of Physics. 
These include the block universe type arguments 
as well arguments motivated by the difficulties in incorporation 
of time in some attempts at formulating Quantum Gravity,
even though in practice Broad's world-view is the dominant one
employed in the Quantum Gravity literature.
We have given a number of reasons why such assumptions 
are problematic as are arguments that employ them. 
These can be summarized by
saying {\it we have time because we shall always be partially ignorant}.
Secondly, solipsist type arguments, according to which only present exist while 
denying the existence of past or future. 
We have argued against such metaphysical arguments, by pointing out that
their implicit denial of history renders them circular and hence inappropriate 
as arguments against the existence of time.

%----------------------------
\section*{Acknowledgments}
%---------------------------

\ni E.A. thanks Julian Barbour, Jeremy Butterfield, Chris Isham 
and Don Page for discussions over the years.
R.T. thanks George Ellis and Tim Clifton for discussions.

%=======================================================================

%_____________________________________________________________________________


\begin{thebibliography}{}
%======================================================================

\footnotesize

\bibitem{PW83} D.N. Page and W.K. Wootters, `Evolution Without Evolution:
Dynamics Described by Stationary Observables', Phys. Rev. {\bf D27}, 2885 (1983).

\bibitem{NSI} S.W. Hawking, `Quantum Cosmology', in {\it Relativity,
Groups and Topology II} ed. B.S. DeWitt and R. Stora (North-Holland, Amsterdam 1984);
%
W.G. Unruh and R.M. Wald, `Time and the Interpretation of Canonical 
Quantum Gravity', Phys. Rev. {\bf D40} 2598 (1989).

\bibitem{B94I}  J.B. Barbour, `The Timelessness of Quantum Gravity: I. The
Evidence from the Classical Theory', Class. Quant. Grav. {\bf 11} 2853 (1994).

\bibitem{B94II} J.B. Barbour, `The Timelessness of Quantum Gravity. II. The Appearance of Dynamics in Static Configurations', 
Class. Quant. Grav. {\bf 11} 2875 (1994).

\bibitem{Page} D.N. Page, `Sensible Quantum Mechanics: Are Probabilities
only in the Mind?', Int. J. Mod. Phys. D5 583 (1996), quant-ph/9507024;
%
`Consciousness and the Quantum', arXiv:1102.5339.

\bibitem{Rec-In-Hist}  M. Gell-Mann and J.B. Hartle, `Decoherence as a
Fundamental Phenomenon in Quantum Dynamics', Phys. Rev. {\bf D47} 3345 (1993).

\bibitem{H99} J.J. Halliwell, `Somewhere in the Universe: 
Where is the Information Stored When Histories Decohere?', 
Phys. Rev. {\bf D60} 105031 (1999), quant-ph/9902008.

\bibitem{Rovelli} C. Rovelli, {\it Quantum Gravity} (Cambridge University Press, Cambridge 2004).

\bibitem{H03} J.J. Halliwell, `The Interpretation of Quantum Cosmology
and the Problem of Time', in {\it The Future of Theoretical Physics and
Cosmology} ed. G.W. Gibbons, E.P.S. Shellard and S.J. Rankin
(Cambridge University Press, Cambridge 2003), arXiv:gr-qc/0208018.

\bibitem{Earman-Block} J. Earman, `Reassessing the Prospects for a Growing Block
Model of the Universe', Int. Stud. Phil. Sci {\bf 22} 135 (2008).

\bibitem{Capetown-Proc} Articles from {\it Do we Need a Physics of Passage?}
Conference being put together by J. N.Y. Acad. Sci,
in particular the ones by Price, Butterfield and Ellis.

\bibitem{Jammerteneity} M. Jammer {\it Concepts of Simultaneity.
From Antiquity to Einstein and Beyond}
(Johns Hopkins University Press, Baltimore 2006).

\bibitem{Broad} C.D. Broad {\it Scientific Thought} (Routledge, London 1923).
 
\bibitem{L} See e.g. {\it The Leibnitz--Clark Correspondence},
ed. H.G. Alexander (Manchester 1956).

\bibitem{Whitrow}  G.J. Whitrow, {\it The Natural Philosophy of Time} 
(Nelson, London 1961). 

\bibitem{Whitrow-89}  G.J. Whitrow, {\it Time in History} (OUP, 1989)

\bibitem{M} E. Mach, {\it Die Mechanik in ihrer Entwickelung,
Historisch-kritisch dargestellt} (J.A. Barth, Leipzig 1883).
%
An English translation is {\it The Science of Mechanics:
A Critical and Historical Account of its Development}
Open Court, La  Salle, Ill. 1960).

\bibitem{Kuchar92} K.V. Kuchar, `Time and Interpretations of Quantum
Gravity', in {\it Proceedings of the 4th Canadian Conference on
General Relativity and Relativistic Astrophysics} ed. G. Kunstatter,
D. Vincent and J. Williams (World Scientific, Singapore, 1992),
reprinted as Int. J. Mod. Phys. Proc. Suppl. {\bf D 20}, 3 (2011).

\bibitem{I93} C.J. Isham, `Canonical Quantum Gravity and the Problem of Time',
In {\it Salamanca 1992, Proceedings, Integrable Systems, Quantum Groups, and 
Quantum Field Theories} 157-287, gr-qc/9210011.

\bibitem{Peres} A. Peres, `Critique of the Wheeler DeWitt equation',
in {\it On Einstein's Path} , Ed. A Harvey
(Springer: Heidelberg): 367-379, 1998 [arXiv:gr-qc/9704061v2]

\bibitem{Bell}  J.S. Bell, `Quantum Mechanics for Cosmologist', 
in {\it Quantum Gravity 2.  A Second Oxford Symposium} ed. 
C.J. Isham, R. Penrose and D.W. Sciama (Clarendon, Oxford, 1981). 

\bibitem{HallHaw} J.J. Halliwell and S.W. Hawking,
`Origin of Structure in the Universe', Phys. Rev. {\bf D31}, 1777 (1985).

\bibitem{Kiefer} C. Kiefer, {\it Quantum Gravity} (Clarendon, Oxford 2004).  

\bibitem{Cartwright} N. Cartwright, {\it How the Laws of Physics Lie}, (OUP, 1983)

\bibitem{HK-RT-1993} H. Kamminga and R. Tavakol, `How Untidy is God's Mind?
A Note on the Dynamical Implications of Nancy Cartwright's Metaphysics',
Brit. J. Phil. Sci., {\bf 44}, 549-553, (1993).

\bibitem{Wald-HE}   R.M. Wald {\it General Relativity} (University of Chicago Press, Chicago 1984);
%
S.W. Hawking and G.F.R. Ellis, {\it The Large-Scale Structure of Space-Time} (Cambridge University Press, Cambridge 1973).  

\bibitem{Tavakol78} R. Tavakol, `Is the Sun Almost Intransitive?'
Nature, {\bf 276}, 802-803 (1978).

\bibitem{Battelle}  J.A. Wheeler, `Superspace and the Nature of Quantum 
Geometrodynamics', in {\it Battelle Rencontres: 1967 Lectures in Mathematics 
and Physics} ed. C. DeWitt and J.A. Wheeler (Benjamin, New York 1968).

\bibitem{Anderson} E. Anderson, `Problem of Time in Quantum Gravity',
Annalen der Physik, {\bf 524} 757 (2012), arXiv:1206.2403;
%
`Problem of Time and Background Independence': the Individual Facets,
arXiv:1409.4117.

\bibitem{Boulware} D. Boulware, `Quantization of Higher Derivative Theories of Gravity', 
in {\it Quantum Theory of Gravity} ed. S. Christensen (Adam Hilger, Bristol, 1984).  

\bibitem{York73} J.W. York Jr., `Mapping onto Solutions of the Gravitational 
Initial Value Problem', J. Math. Phys. {\bf 13} 125 (1972).  

\bibitem{Mott}   N. Mott, `The Wave Mechanics of $\alpha$-Ray Tracks', Proc. Roy. Soc. Lon. {\bf A126} 79 (1929).   
 
\bibitem{Records} E. Anderson, `Records Theory', Int. J. Mod. Phys.
{\bf D18} 635 (2009), arXiv:0709.1892;
%
`Kendall's Shape Statistics as a Classical Realization of
Barbour-type Timeless Records Theory approach to Quantum Gravity', Accepted by SHPMP, arXiv:1307.1923. 

\bibitem{Ellis-2012}  G. F. R. Ellis and R. Goswami,
`Space Time and the passage of Time', arXiv:1208.2611. 

\bibitem{Ellis-2010} G. F. R. Ellis and T. Rothman,
`Time and Spacetime: The Crystallizing Block Universe',
Int. J. Theor. Phys.  {\bf 49}, 988 (2010) [arXiv:0912.0808 [quant-ph]]. 

\bibitem{JZ} E. Joos and H.D. Zeh, `The Emergence of Classical Properties 
Through Interaction with the Environment', Z. Phys. {\bf B59} 223 (1985).

\bibitem{PAOT}  D.N. Page, in {\it Physical Origins of Time Asymmetry} ed. J.J. Halliwell, J. Perez-Mercader and W.H. Zurek (Cambridge University Press, Cambridge, 1994).  

\bibitem{IGUS}                J.B. Hartle, {The Physics of 'Now'}, Am. J. Phys. {\bf 73} 101 (2005), gr-qc/0403001. 

\bibitem{EOT} J.B. Barbour, {\it The End of Time} (Oxford University Press, Oxford 1999).

\bibitem{Sorkin}  R.D. Sorkin, `Causal Sets: Discrete Gravity'
(Notes for the Valdivia Summer School) gr-qc/0309009.

\bibitem{Dowker} H.F. Dowker, `The Birth of Spacetime Atoms as the 
Passage of Time', arXiv:1405.3492.

\bibitem{HP11} V. Husain and T. Pawlowski, `Time and a Physical Hamiltonian for Quantum Gravity', Phys. Rev. Lett. {\bf 108} 141301 (2012), arXiv:1108.1145; 
%
`Dust Reference Frame in Quantum Cosmology', Class. Quant. Grav. {\bf 28} 225014 (2011), arXiv:1108.1147. 

\bibitem{BSW} R.F. Baierlein, D.H. Sharp and J.A. Wheeler,
`Three-Dimensional Geometry as Carrier of Information about Time',
Phys. Rev. {\bf 126} 1864 (1962).

\bibitem{AM13} E. Anderson and F. Mercati, `Classical Machian Resolution
of the Spacetime Construction Problem', arXiv:1311.6541.
%_____________________________________________________________________________
\end{thebibliography}
\end{document}